# The Curious Association Between Hadley Circulation Intensity and the Meridional Distribution of Tropical Cyclones in the Eastern North Pacific

Joshua Studholme[1][*] and Sergey Gulev[1,2]

1. *Shirshov Institute of Oceanology, Russian Academy of Sciences*

2. *Lomonosov Moscow State University*

*Moscow, Russian Federation*

An English language biography of Alexander Obukhov can be read here:
https://sci-hub.tw/https://link.springer.com/article/10.1007%2FBF00122458

Date: 19th Jul. 2018

---

*\* Corresponding author address: Joshua Studholme, Shirshov Institute of Oceanology, RAS. 36 Nakhimovsky ave. 117997. Moscow. Russian Federation.*

*Email: josh.studholme@gmail.com*

ABSTRACT

It has recently been shown that an interesting inverse relationship exists between the strength of the overturning in the regional boreal Hadley circulation and tropical cyclone genesis and lifetime maximum intensity latitudes in the eastern North Pacific. We show here that this curious result can be understood as the outcome of an equatorward shift in the Intertropical Convergence Zone (ITCZ) and a moderate reduction of tropical vertical shear. These two factors are not necessarily concomitant. The magnitude of the vertical shear change is low enough to suggest that the primary physical mechanism behind this inverse relationship is the equatorward shift in the ITCZ. Since a significant proportion of tropical cyclone genesis results from aggregated convective cells forming into a coherent convective vortex and being shed from the ITCZ, this shift is apparently sufficient to explain the observed association. The equatorward shift in the ITCZ is potentially the result of a relative warming of the Southern Hemisphere in the boreal tropical cyclone season. This more equatorward ITCZ, located over warmer surface waters, presumably explains the more intense Hadley circulation, despite a reduction of the large-scale boreal meridional temperature gradient. Although, these links require further research.

## 1. Introduction

Recent work looking at any concurrencies between seasonal-mean changes to regional Hadley circulation (HC) and the meridional distribution of tropical cyclones (TCs) revealed an interesting inverse association between the genesis and lifetime maximum intensity (LMI) latitudes of TCs in the eastern North Pacific (Studholme and Gulev, 2018; hereafter SG18). This result is not only particularly curious because of the high degree of shared detrended covariance between these two phenomena, the coefficient of determination being approximately $1/3^{rd}$ of the interannual variability, but also since it is not found in any other TC-spawning ocean basin. While comparable magnitude concurrency was found between HC terminus latitudes (alternatively referred to as HC extent), the actual strength of the overturning itself was found to be relevant only in this specific ocean basin.

Relatively robust environmental controls on TC distribution have been both quantitively (e.g. Emanuel, 1988, 1995; Bister and Emanuel, 1998) and qualitatively (e.g. Emanuel and Nolan, 2004; Tippett et al., 2011) well established, although significant issues remain in implementing and interpreting these, particularly with regards to the qualitative indices and how they relate to tropical cyclogenesis (e.g. Tory et al., 2018). At the same time, over the last decade or so it has been shown that the first-principle theoretical model for an axisymmetric HC at the nearly inviscid limit derived in and around the 1980s (Schneider, 1977; Held and Hou, 1980; Linzen and Hou, 1988) does not agree with eddy resolving numerical model simulations (e.g. Schneider, 2006). The reason for this may be understood by considering the mean zonal momentum balance in the upper troposphere above the Hadley cells with regards to the local Rossby number (Ro $= -\bar{\zeta}/f$; Levine and Schneider, 2011). Eddy momentum fluxes may only be ignored at the limit Ro $\rightarrow 1$. In nature, the Ro in upper branches of the HC is ~0.4 and thus eddy momentum fluxes are non-negligible and we lack an appropriate theoretical model within to understand HC dynamics.

Reconciling these vast open fundamental theoretical questions is well beyond the scope of this short research note. We must also be careful not to project causality upon this curious local association in the eastern North Pacific. The aim therefore of this paper is to present and provide a discussion of the large-scale zonal-mean changes in the atmospheric-oceanic state associated with a weakening of the local HC and an equatorward shift in tropical cyclogenesis and LMI latitudes. Data and methods are given in the subsequent section before the presentation of the results in section 3. The manuscript ends with a summary and terse discussion.

## 2. Data and Methodology

Here we use data related to TCs and the large-scale ocean-atmospheric state. The TC record is taken from the International Best Track Archive for Climate Stewardship (IBTrACS), version 03r10 (Knapp et al., 2010). We use the same data processing procedure as in SG18 which is largely based on the general principles discussed in Schreck et al. (2014). For brevity here we only reproduce results using the National Hurricane Center (NHC) and Joint Typhoon Warning Center (JTWC) aggregation. Results have been verified against an alternate dataset based on the IBTrACS-WMO homogenization product and the conclusions discussed here are not affected. The time period under analysis is 1981 to 2016; matching the period used in SG18.

The meridional overturning streamfunction, $\psi$, is derived from the divergent winds computed from the ERA-Interim reanalysis (Dee et al., 2011). The same reanalysis is used for data on sea surface temperature (SST), tropospheric temperature, relative humidity, vertical shear (200 – 850 hPa wind vector difference), and cloud liquid and ice water content. These data were downloaded at ¾° x ¾° spatial resolution and monthly-mean temporal resolution. Zonal averaging is done over the eastern North Pacific region between 105°W-150°W. Small (±10°) deviations either side of these boundaries were found to not alter the results. Temporal averaging is done over the July-August-September-October (JASO) period, defined as the boreal TC season. Timeseries for the divergent overturning intensity in the boreal Hadley cell is computed using the method detailed in SG18. Briefly, the intensity of the cell is taken as the vertically averaged maximum value of the JASO-mean $\psi$ between 900 and 200 hPa within the cell. The Niño3 index (SST anomaly over 5°S-5°N, 150°W-90°W) is computed from ERA-Interim. What we refer to here as 'strong El Niño' and 'strong La Niña' years are those that occur when the JASO-mean Niño3 index is within the 90[th] and 10[th] percentiles respectively. This makes the strong El Niño years 1982, 1987, 1997, and 2015.

The strong La Niña years are 1988, 1999, 2007 and 2010.

## 3. Results

### a. *Eastern North Pacific Hadley Circulation and mean tropical cyclone latitudes*

The robust inverse relationship with the strength of the JASO boreal HC and the latitude of TC genesis is shown in figure 1. Understanding the physical dynamics behind this relation is the subject of the present work. The Pearson product moment coefficient using mean genesis latitudes for TCs is -0.60 (two-tailed *p*-value <0.001) including all years, while excluding ENSO years (all shaded points) the same coefficient is -0.51 (*p*-value <0.01). Thus, this relationship is robust when ENSO is accounted for despite the fact that ENSO is the strongest modulator of HC intensity and one of the strongest modulators of TC latitudes in this region. While here we show only the mean TC genesis latitudes, this relationship is observable across the entire distribution of genesis and LMI latitudes. For example, using the equatorward side of TC genesis latitudes (P(10)) these correlation values are -0.65 (*p*-value $<1\times10^{-5}$) including all points and -0.55 (*p*-value <0.001) when strong ENSO years are excluded.

Annual-mean and seasonal-mean (JASO) HC intensity was studied in SG18. It was shown that there is generally sound correspondence between reanalysis products (ERA-Interim, JRA55 and MERRA2) in the eastern North Pacific. Having said that, JRA55 diverged from Interim and MERRA2 around the turn of the millennium when it produces consistently higher magnitude HC intensity than the other two products. Throughout this time, the interannual variability however is consistent. SG18 found no strong trend in eastern North Pacific HC intensity across all reanalyzes. Meanwhile, it was shown that the meridional distribution TC genesis latitudes in this basin have shifted wholesale towards the pole with a rate of approximately $0.3°$ latitude decade$^{-1}$. No trend was found in LMI latitudes there. Thus, the association between local Hadley overturning and TC latitudes persists after detrending and it is observed consistently outside of ENSO years.

*b. Local Hadley Circulation characteristics and interseasonal variability*

HC is climatologically very weak and shallow in the eastern North Pacific (figure 2a). It is the region of weakest HC of all ocean basins that experience TCs. In fact, it also experiences no real zonal overturning (i.e. local Walker circulation) to speak of either and it is also not exposed to a strong subtropical jet like its counterparts (e.g. Schwendike et al. 2014). This latter point is worth noting in particular since SG18 hypothesized that the local zonal atmospheric flow and the Walker circulation in the eastern Pacific might be the root cause of this curious association. This suggestion may now be discarded.

Mean tropical cyclogenesis in the basin occurs at 13.7°N which is just north of the climatological Intertropical Convergence Zone (ITCZ; figure 2a). Local overturning in the boreal summer is characterized by the strong protrusion of the austral Hadley cell into the Northern Hemisphere. This austral cell is much deep and stronger than its boreal counterpart, and much more resembles the archetypical Hadley cell, seen for example in the global zonal-mean (e.g. Nguyen et al., 2013). The boreal cell here is weak, shallow and narrow.

Interannual variability of local HC is characterized along three primary axes: the depths and overturning intensities of both Hadley cells and the ITCZ latitude (figure 2b). The strongest zone of interannual variability in $\bar{\psi}_{JASO}$ is around 8°N, just shy of the ITCZ latitude (as inferred from figure 2a). Magnitudes of STD here are around 50 x $10^9$ kg s$^{-1}$, around 25% of the climatological mean. Since a large contribution of total tropical cyclogenesis is provided by convective clusters spawned from the ITCZ (e.g. Fedorov et al., 2018), it would seem reasonable that this therefore might be behind the observed concurrent equatorward shift of TC latitudes and HC intensity. The reasoning being that a stronger boreal Hadley cell is associated with an equatorward shift in the ITCZ. This of course relies upon the robustness of the metric for HC intensity, and HC diagnostic algorithms applied to the divergent meridional overturning are notoriously tricky (e.g. Nguyen et al., 2018; SG18). Although luckily, these issues mostly relate only to the determination of HC terminus latitudes and not to overturning intensity.

*c. Associated large-scale ocean-atmosphere state*

Any directly causal relation between Hadley overturning intensity is unlikely, but presumable possible thus we must take open view of possible physical linkages behind this observed inverse relationship. Such a causal link might for example occur via HC's influence on vertical wind shear. SG18 showed that stronger boreal Hadley cell intensity can be linearly related to a sharp increase in ($0.15$ ms$^{-1}$ [$1 \times 10^9$ kg s$^{-1}$]$^{-1}$) in JASO vertical wind shear in a region around 140°W 25°N and a similarly sharp decrease in a band tightly hugging to the Central American coastline. The aim of the present analysis is to understand whether this indeed corresponds directly solely to the strength of the overturning itself or is rather a manifestation of a local shift in the ITCZ latitude. We attempt to do this by taking composites of the 75[th] and 25[th] percentile years in the boreal Hadley cell's intensity (figure 3). These years are 1982, 1986, 1987, 1994, 1997, 2002, 2009, 2014, 2015 for P(75) and 1988, 1995, 1996, 1998, 1999, 2007, 2008, 2010, 2011 for P(25).

The years of the most intense HC in the Northern Hemisphere demonstrate a strong association between the strength of the cell and its depth (figure 3a). Indeed, the weakest years (figure 3b) also show horizonal confinement of the cell, the extent diagnostic delta between the two composites is 5.6° latitude. However, the delta between the two composites (figure 3c) reduces the number of likely candidates for the association to the circulation depth and the ITCZ latitude. The strongest difference is at the near-surface around 8°N which again presumably corresponds to the location of the ITCZ while the high magnitude delta between 15 and 20°N at 500 hPa, right above the climatological maximum overturning cell (figure 2a), highlights that a stronger HC is a deeper HC, however unsurprising as this result may be.

We compute the corresponding zonal-mean composites for a number of atmospheric-oceanic state variables relevant to both HC and TCs. We show these as biases from the climatological mean computed over the entire period (1981-2016; figure 4). Years of intense HC (red curve) in the eastern North Pacific occur with on average a >1K equatorial SST anomaly, and a >2K difference relative to years of the weakest boreal overturning (blue curve; figure 4a). This equatorial warming is over x5 times the magnitude of a more general warming in the tropical and subtropical ocean of ~0.2K that slowly tappers off up to the midlatitudes. This warming in the years of intense HC is mirrored for years of the weakest HC. This result agrees with the linear regression analysis of SG18. The ocean surface warming is associated with a bulk warming (~0.75K) in the column-mean tropospheric temperature in the tropics and subtropics but cooling of comparable magnitude in the midlatitudes (figure 4b). The latitude of maximum temperature anomaly is 20°S, i.e. well offset from the maximum SST anomaly located at the Equator. These temperature changes are associated with a moistening of the mid-troposphere (specifically at 500hPa) by up to 10% relative to climatology (figure 4c).

The perturbation of the climatological tropospheric meridional temperature gradient (figure 4d) is associated with strong changes to vertical shear (figure 4f). In both hemispheres, years with stronger boreal HC is associated with reduced tropical shear and increased tropical shear. In the Northern Hemisphere, at TC latitudes these changes are ~2.5 ms$^{-2}$. This change in the shear environment seems to follow directly from the changes in the meridional temperature gradient. Through the thermal wind relation, i.e. $\partial u/\partial z \propto \partial T/\partial y$ (using standard geophysical notation; e.g. Holton and Hakim, 2013), a relaxation of the tropospheric meridional temperature will invoke a reduction in the vertical shear (the inverse being true also). In the tropics, either side of the climatological ITCZ, years with a more intense boreal HC are associated with a flatter tropical temperature gradient and a steeper subtropical gradient (figure 4d). These changes are again mirrored in years with less intense boreal HC. These gradient changes correspond exactly to the observed vertical shear differences (figure 4f).

The years of more intense boreal HC are also associated with a clear equatorward shift in the ITCZ (figure 4e). We show column integrated cloud liquid and ice water here as a proxy for tropical deep moist convection since this is maximal for the tropics in the ITCZ. The difference in the anomalies in the concentration of cloud water between the two composites is >0.05 kg m$^{-2}$ and is approximately symmetric around the climatological ITCZ suggesting the shift is unambiguous.

## 4. Summary and discussion

Here we have shown that the curious inverse relationship between HC intensity, and tropical cyclogenesis and LMI latitudes in the eastern North Pacific can be understood to be the result an equatorward shift in the ITCZ and a moderate reduction of tropical vertical shear. The magnitude of the vertical shear changes is low enough to suggest that the primary physical mechanism behind this inverse relationship is the equatorward shift in the ITCZ. Since a significant proportion of tropical cyclone genesis results from convective vortices being shed from the ITCZ this explains the observed association.

It follows therefore that a more intense boreal Hadley cell is associated with a more equatorward JASO ITCZ. This warrants brief comment here but it largely beyond the scope of the present objective. In the absence of robust physical theory for the ITCZ proper, typically the observation that the ITCZ latitude ($\varphi_{ITCZ}$) is proportional to the cross-equatorial atmospheric energy transport ($H$) is used to understand ITCZ shifts (e.g. Schneider et al., 2014). This is given by the difference between the two hemispheres' ($\Delta$) respective atmospheric energy influx by shortwave ($S$) and longwave ($L$) radiative fluxes at the top of the atmosphere, and aggregate surface radiative and turbulent fluxes ($O$) (Kang et al., 2009). Thus specifically:

$$\varphi_{ITCZ} \propto H \propto \Delta(S - L - O),$$

This is a reflection that any hemispheric asymmetry in energy input to the atmosphere may be efficiently balanced by an anomalous cross-equatorial HC exporting dry static energy from the heated hemisphere into the cooled hemisphere via its upper-level flow and importing atmospheric moisture in its low levels via the corresponding return flow (e.g. Voigt et al., 2017). Presumably then, the equatorward shift in the ITCZ during the years of most intense boreal HC could be reflection of the relatively warmer southern hemisphere and potentially also cloud radiative responses associated with the increase in boreal humidity. An equatorward shift in the ITCZ could then act to intensify HC since, although the large-scale meridional SST gradient is reduced, the ITCZ (which houses the convective updrafts in part driving HC) has been shifted to be over warmer surface waters.

***Acknowledgements.*** This research was supported by the Russian Ministry of Education and Science through agreement 14.616.21.0075, project ID RFMEFI61617X0075. We thank the ECMWF, and NOAA for releasing their data to the public and the open-source Python and data-analysis community.

## LIST OF FIGURES

**Fig. 1.** Figure 1: JASO seasonal-mean HC intensity from the ERA-Interim reanalysis against TC mean genesis latitudes (red) from the NHC-JTWC observational record, the eastern North Pacific. Using the Niño3 index, strong El Niño years are shaded red and La Niña years in blue. The Pearson product moment coefficient is -0.60 (two-tailed *p*-value <0.001) including all years. Excluding from ENSO years (all shaded points) it is -0.51 (*p*-value <0.01).

**Fig. 2.** Climatological mean (a) and STD (b) for the divergent overturning streamfunction zonal-mean in the eastern North Pacific JASO. White vertical line marks the latitude of the equator. Red and blue vertical lines mark the latitudes of the Hadley cell overturning maxima and termini respectively. Units are $1 \times 10^9$ kg s$^{-1}$.

**Fig. 3.** Eastern North Pacific divergent meridional overturning composites based on the boreal Hadley cell's most [ P(75)] ; (a)] and least [ P(25) ; (b)] intense years and the difference between the two (c). Only when differences between the composite means are robust at the 5% level based on a two-sided Students' *t*-test are values included in shading in (c). Units are $1 \times 10^9$ kg s$^{-1}$.

**Fig. 4.** Zonal-mean large-scale state biases relative to climatology in composites' mean years of most (red curves) and least (blue curves) intense boreal Hadley circulation.

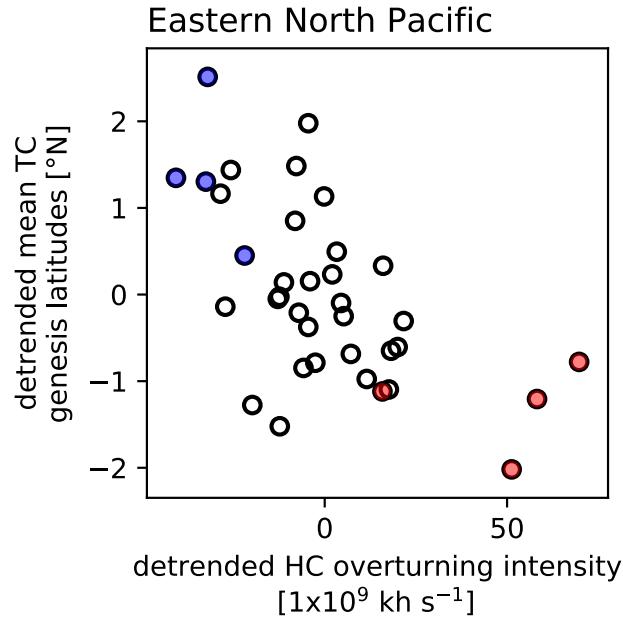

Figure 1: JASO seasonal-mean HC intensity from the ERA-Interim reanalysis against TC mean genesis latitudes from the NHC-JTWC observational record, the eastern North Pacific. Using the Niño3 index, strong El Niño years are shaded red and La Niña years in blue. The Pearson product moment coefficient is -0.60 (two-tailed *p*-value <0.001) including all years. Excluding all ENSO years (all shaded points) it is -0.51 (*p*-value <0.01).

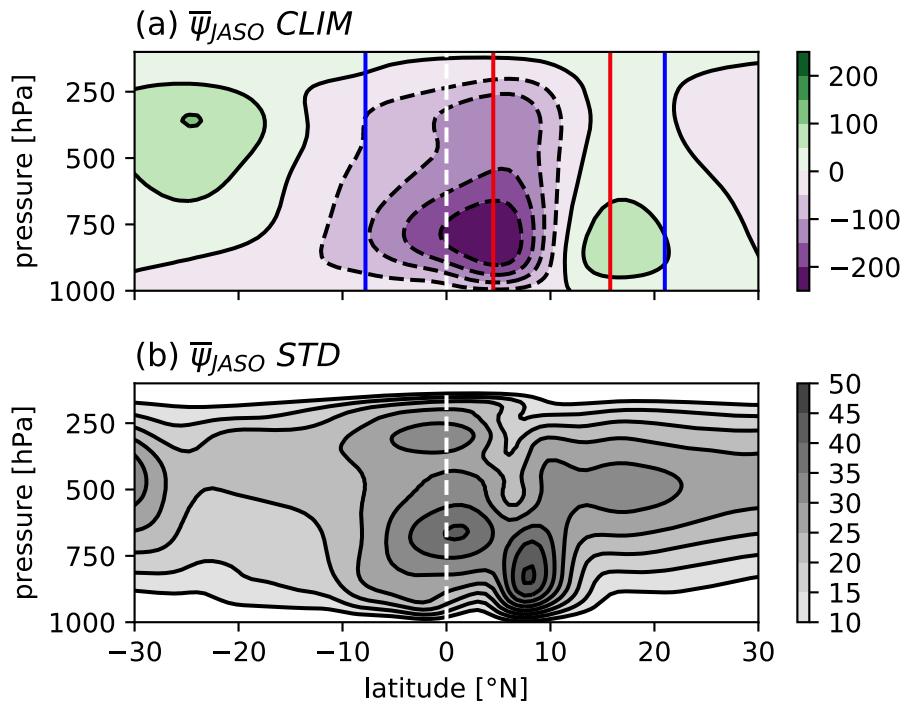

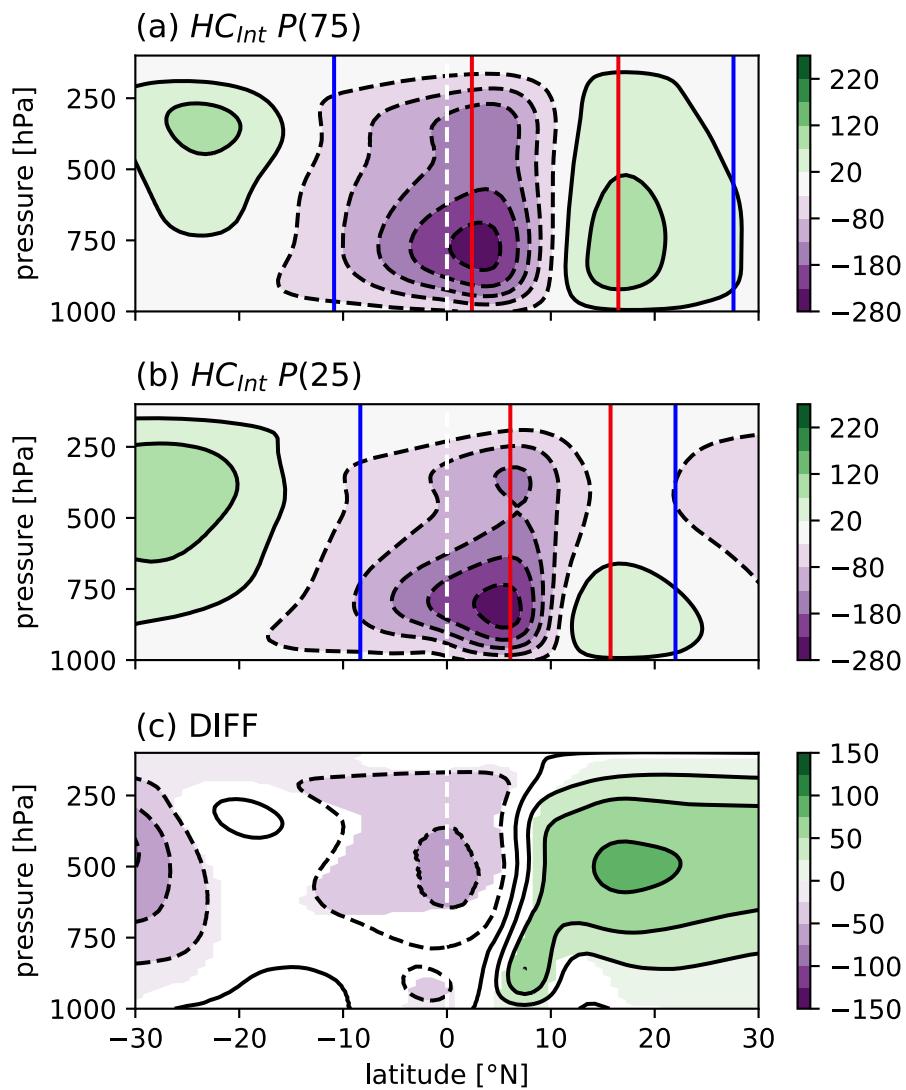

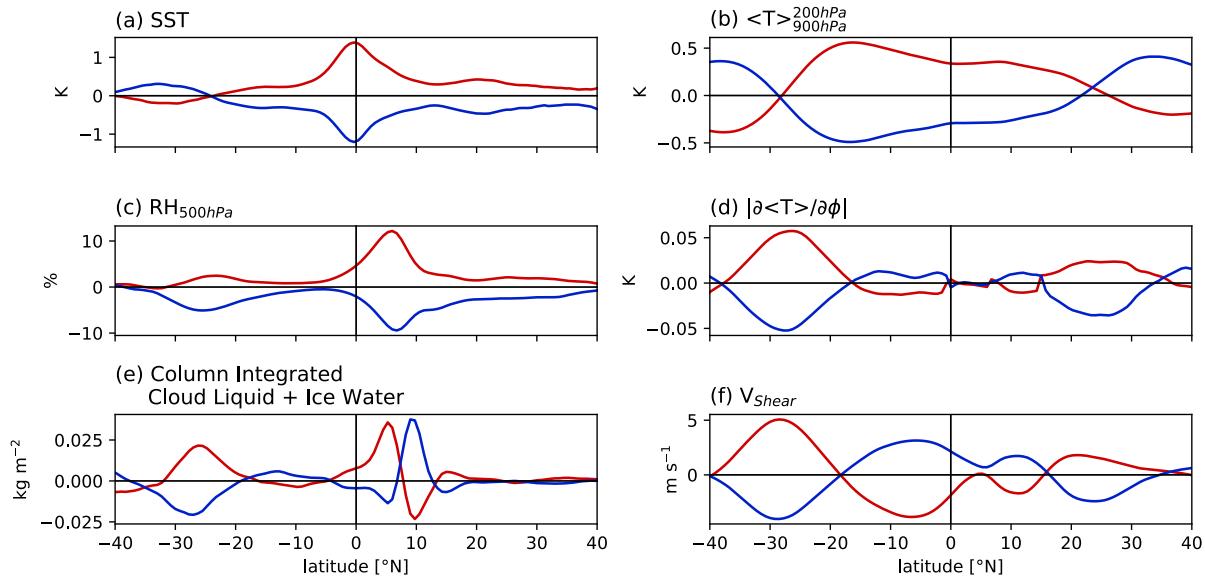